\documentclass{article}
\begin{document}

\centerline{\bf NONCOMMUTATIVE QUANTUM MECHANICS AND GEOMETRY}
\centerline{\bf FROM THE QUANTIZATION IN C-SPACES}
\centerline{Carlos Castro}
\bigskip
\centerline{Center for Theoretical Studies of Physical Systems}
\centerline{Clark Atlanta University, Atlanta GA. 3031}
\centerline{June 2002}
\bigskip
\centerline{\bf Abstract}
\bigskip

Four years ago the Extended Scale Relativity (ESR) theory
in C-spaces (Clifford manifolds) was proposed as the plausible physical
foundations of string theory. In such theory the speed of light and 
the
minimum Planck scale are the two universal invariants. All the 
dimensions
of a C-space can be treated on equal footing by implementing the 
holographic
principle associated with a nested family of p-loops of various
dimensionalities. This is achieved by using polyvector valued 
coordinates in
C-spaces that encode in one stroke points,
lines, areas, volumes,... We review the derivation of
the minimal length/time string/brane uncertainty relations and the 
maximum
Planck temperature thermodynamical uncertainty relation. The 
Weyl-Heisenberg
algebra in C-spaces is constructed which $induces$ a Noncommutative
Geometric structure in the $ X^A $ coordinates. Hence quantization in
C-spaces involves in a natural fashion a
Noncommutative Quantum Mechanics and Field Theory rather than being
postulated ad-hoc. A
QFT in C-spaces may very likely involve (Braided Hopf) Quantum 
Clifford
algebras and generalized Moyal-like star products associated with
multisymplectic geometry.
\bigskip

\centerline {\bf Introduction}
\bigskip
In recent years we have argued that the underlying fundamental physical
principle
behind string theory, not unlike the principle of equivalence and
general covariance in Einstein's general relativity, might well be
related
to the existence of an invariant minimal length scale (Planck scale)
attainable in nature. A scale
relativistic theory involving spacetime $resolultions$ was developed
long ago by Nottale where the Planck scale was postulated as the 
minimum
observer independent invariant resolution [1] in Nature. Since
``points'' cannot be observed physically with an ultimate resolution, 
they
are
fuzzy and smeared out into fuzzy balls of Planck radius of arbitrary
dimension. For this
reason one must construct a theory that includes all dimensions (and
signatures) on the equal footing. Becuase the notion of dimension is a
topological invariant, and the concept of a fixed dimension is lost 
due to
the fuzzy nature of points, dimensions are resolution-dependent, one 
must
also include a theory with $ all $ topologies as well. It is our 
belief
that
this may lead to the proper formulation of string and M theory.

In [2] we applied this Extended Scale Relativity principle to the 
quantum
mechanics of $p$-branes which led to the construction of C-space (a
dimension $category$) where all $p$-branes were taken to be on the same
footing; i.e. transformations in C-space reshuffled a string history 
for a
five-brane history, a membrane history for a string history, for 
example.
It turned out that Clifford algebras contained the appropriate
algebro-geometric features to implement this principle of 
polydimensional
transformations [3, 4, 5].

Clifford algebras have been a very useful tool for a description of 
geometry
and physics [4, 5, 6, 7, 8]. In [3,5] it was proposed that every
physical quantity is in fact a $ polyvector$, that is, a Clifford 
number
or a Clifford aggregate. Also, spinors are the members of left or 
right
minimal ideals of Clifford algebra, which may provide the framework 
for a
deeper understanding of sypersymmetries, i.e., the transformations
relating
bosons and fermions.
The Fock-Stueckelberg theory of relativistic particle [4] can be 
embedded
in the Clifford algebra of spacetime [3]. Many important
aspects of Clifford algebra are described in [3,5,6,7,8]

Using these methods the bosonic
$p$-brane propagator, in the quenched minisuperspace approximation, was
constructed in [9]; the logarithmic corrections to the black hole 
entropy
based on the geometry of Clifford space (in
short $C$-space) were obtained in [12]; the action for higher 
derivative
gravity with torsion from the geometry of C-spaces and how the 
Conformal
agebra of spacetime emerges from the Clifford algebra was performed in
[11]; the resolution of the ordering ambiguities of QFT in curved 
spaces was
resolved by [3].

In this new physical theory the arena for physics is no longer the 
ordinary
spacetime, but a more general manifold of Clifford algebra valued
objects, polyvectors. Such a manifold has been called a
pan-dimensional continuum [5] or $C$-space [2]. The
latter describes on a unified basis the objects of various
dimensionality: not only points, but also closed lines, surfaces,
volumes,.., called 0-loops (points), 1-loops (closed strings) 2-loops
(closed membarnes), 3-loops, etc.. It is a sort
of a $dimension$ category, where the role of functorial maps is
played by C-space transformations which reshuffles a $p$-brane history 
for a
$p'$-brane history or a mixture of all of them, for example.
The above geometric objects may
be considered as to corresponding to the well-known physical
objects, namely closed $p$-branes. Technically those
transformations in C-space that reshuffle objects of different 
dimensions
are generalizations of the ordinary Lorentz transformations to 
$C$-space. In
that sense, the C-space is roughly speaking a sort of generalized
Penrose-Twistor space from which the ordinary spacetime is a $derived$
concept.

In [2] we derived the minimal length uncertainty relations as well as 
the
full blown uncertainty relations due to the contributions of $all$
branes
of $every $ dimensionality, ranging from $ p = 0 $ all the way to $ p =
\infty$.
In [14] we extended
this derivation to include the $maximum$ Planck Temperature condition 
which
recently has been recast into a maximum temperature thermodynamical
uncertainty relation involving the internal Energy, temperature and 
the
Boltzmann constant [19].

In section {\bf I } we will review the Extended Relativity in C-spaces 
and
the explicit derivation, from first principles, of all the generalized
minimal length-time (and maximum temperature) uncertainty relations 
based
on the effective-running Planck ``constant'' $ \hbar $ (energy
dependent) emerging from the geometry of C-spaces.

In section {\bf II } we proceed with the canonical quantization
in C-spaces and construct the Weyl-Heisenberg algebra. We show 
rigorously
how the Noncommutative Geometry of the C-space coordinates $X$ is a 
direct
$consequence$ of the Weyl-Heisenberg algebra in C-spaces in
contradisctinction with the $ordinary$ phase space commutators
$ [ x^i, p^j ] = i \hbar \delta^{ij} $
which imply that $[ x^i, x^j ] = 0$ as a
result of the Jacobi indentities.

In C-space this will $no$ longer be the case due to the highly 
nontrivial
Weyl-Heisenberg algebra. This is the main result of this work: 
Canonical
Quantization in C-spaces automatically yields a Noncommutative 
Geometric
structure for the coordinates $X$ and hence it involves a 
Noncommutative QM.
It is unnecessary to put in by hand the noncommutativity of the
coordinates in terms of a length scale (like the Planck scale) like
Snyder did long ago [22].
The mere quantization in C-spaces $induces$ the Noncommutativity of
coordinates. We believe this is an important result deeply ingrained in 
the
Extended Scale Relativistic structure of C-spaces.

We finalize by discussing how one can extract an effective $ \hbar$, 
which
is a function of the Lorentz invariant quantity $ p^2 = p_\mu p^\mu $,
from the Weyl-Heisenberg algebra in C-space. Effective $ \hbar$ of 
this
sort are the ones which furnish the minimal length/time string/brane
uncertainty relations. We conclude with some remarks pertaining
multisymplectic geometry and QFT in C-spaces based on Quantum Clifford
Algebras and Braided Hopf Quantum Cliford algebras to study 
q-deformations
of C-spaces.

A Moyal star product construction deserves further study as well. Since
C-spaces involves the physics of all $ p$-loops it is warranted to use
methods of multisymplectic mechanics since phase spaces in C-spaces 
involve
antisymmetric tensors of arbitrary rank. Nambu-Poison QM seems to be 
the
most appropriate one to study C-space QM. In particular the use of the
Zariski star product deformations vz the Moyal one will be welcome 
[24].

\bigskip

\centerline {\bf I. Extending Relativity from Minkowski spacetime to
$C$-space}
\bigskip
We embark into the extended relativity theory in
C-spaces by a natural generalization of the notion of a space-
time interval in Minkwoski space to C-space:

$$ dX^2 = d\Omega^2 + dx_\mu dx^\mu + d x_{\mu\nu} dx^{\mu\nu} + ...
\eqno ( 1 ) $$ The Clifford valued poly-vector:
$$ X = \Omega I + x^\mu \gamma_\mu +
x^{\mu\nu} \gamma_\mu \wedge \gamma_\nu + ... \eqno ( 2 ) $$
denotes the position in a manifold, called Clifford space or
$C$-space. The coordinates $ x^{\mu\nu}, ....$
are the holographic areas, volumes, ...shadows or proyections of the 
nested
family of $ p$-loops onto the embedding spacetime coordinate
planes/hyperplanes. Since the Planck scale is given by:
$ \Lambda = ( G_N )^{ 1/ (D - 2) }$ in $ D$ dimensions, in units of 
$
\hbar = c = 1 $, and since we wish to have a universal value for the 
minimum
distance in $all$ dimensions, we will set
$ \Lambda = G_N = 1 $, which is the consistent value in $all$ 
dimensions
ranging from
$ D = 2 $ all the way to $ D = \infty $. The ESR theory admits 
naturally
this sytem of units, of setting all fundamental constants to $unity$,
including Boltzmann constant and the Planck Temperature.

If we take differential $ d X$ of $X$ and compute the
scalar product $ d X * dX$ we obtain:

$$ d \Sigma^2 = ( d \Omega)^2 + \Lambda^{ 2D -- 2 } dx_\mu dx^\mu 
+
\Lambda^{ 2D -- 4 } d x_{\mu\nu} dx^{\mu\nu } + ..\eqno(3) $$

Here we have introduced the Planck scale $\Lambda$ since a
length parameter is needed in order to tie objects of different
dimensionality together: 0-loops, 1-loops,..., $p$-loops. Einstein
introduced the speed of light as a universal absolute invariant in
order to ``unite'' space with time (to match units) in the
Minkwoski space interval:
$$ d s^2 = c^2 d t^2 - d x_i d x^i . \eqno ( 4 ) $$
A similar unification is needed here to ``unite'' objects of
different dimensions, such as $x^\mu$, $x^{\mu \nu}$, etc... The
Planck scale then emerges as another universal invariant in
constructing an extended scale relativity theory in C-spaces [2].

To continue along the same path, we consider the analog of Lorentz
transformations in C-spaces which transform a poly-vector
$X$ into another poly-vector $X'$ given by $ X' = R X R^{-1} $
with
$$ R = exp~ [ i ( \theta I + \theta^\mu \gamma_\mu +
\theta^{\mu_1 \mu_2 } \gamma_{\mu_1} \wedge \gamma_{\mu_2 } .....)
] . \eqno ( 5 ) $$ and

$$ R^{-1} = exp~[ - i ( \theta I + \theta^\nu \gamma_\nu +
\theta^{\nu_1
\nu_2 } \gamma_{\nu_1} \wedge \gamma_{\nu_2 } .....) ] .\eqno ( 6
) $$ where the theta parameters:

$$ \theta; \theta^\mu; \theta^{\mu\nu}; ... . \eqno ( 7 ) $$
are the C-space version of the Lorentz rotations/boosts
parameters.

Since a Clifford algebra admits a matrix representation, one can
write the norm of a poly-vectors in terms of the trace operation
as: $ || X ||^2 = {Trace} ~ X^2 $ Hence under C-space Lorentz
transformation the norms of poly-vectors behave like follows:
$$ Trace~ {X'}^2 = Trace ~[ R X^2 R^{-1} ] = Trace ~
[ R R^{-1} X^2 ] = Trace~ X^2 . \eqno ( 8 ) $$
These norms are invariant under C-space Lorentz transformations
due to the cyclic property of the trace operation and $ R R^{-1} =
1 $.

\bigskip

\centerline {\bf 1.2 Planck scale as the minimum invariant in
Extended Scale Relativity }

\bigskip

Long time ago L.Nottale proposed to view the Planck scale as
the absolute minimum invariant (observer independent) scale in
Nature in his formulation of scale relativity [1]. We can apply
this idea to C-spaces by choosing the correct analog of the
Minkowski signature:

$$ || d X || ^2 = d \Sigma^2 = ( d \Omega )^2 [ 1 - \Lambda^{ 2D - 2 }
{ ( dx_\mu)^2 \over ( d \Omega)^2 } - \Lambda^{ 2D - 4 } {
(dx_{\mu\nu} )^2 \over ( d \Omega)^2 } - \Lambda^{ 2D - 6 } {
(dx_{\mu\nu\rho} )^2 \over ( d \Omega)^2 } - ..] $$

$$ || d X || ^2 = d \Sigma^2 = ( d \Omega )^2
[ 1 - ({ \Lambda \over \lambda_1} )^{ 2D - 2 } - ({ \Lambda \over
\lambda_2} )^{ 2D - 4 } - ({ \Lambda \over \lambda_3} )^{ 2D -
6 } - ...] . \eqno ( 9 ) $$

where the sequence of $variable$ scales $ \lambda_1, \lambda_2,
\lambda_3, ....$ are related to the generalized (holographic)
velocities defined as follows:

$$ { ( dx_\mu)^2 \over ( d \Omega)^2 } \equiv ( V_1)^2 =
( { 1 \over \lambda_1 } )^{ 2D - 2 } . $$

$$ { ( dx_{\mu\nu} )^2 \over ( d \Omega)^2 } \equiv ( V_2)^2 =
( { 1\over \lambda_2 } )^{ 2D - 4 } . $$

$$ { ( dx_{\mu\nu\rho} )^2 \over ( d \Omega)^2 } \equiv ( V_3)^2 =
({ 1 \over \lambda_3 } )^{ 2D - 6 } . . \eqno ( 10 ). $$

. . .

It is clear now that if $ || dX ||^2 \ge 0 $ then the sequence of
$variable$ lengths $ \lambda_n$ cannot be $smaller$ than the
Planck scale $ \Lambda $. This is analogous to a situation with
the Minkoswki interval:

$$ ds^2 = c^2 dt^2 [ 1 - {v^2\over c^2 } ] . \eqno ( 11 ) $$

when it is $ \ge 0 $ if, and only if, the velocity $ v$ does not
exceed the speed of light. If any of the $\lambda_n$ were smaller
than the Planck scale the C-space interval will become
tachyonic-like $ d \Sigma^2 < 0 $. Photons in C-space are
$tensionless$ branes/loops. Quite analogously one can interpret
the Planck scale as the postulated minimum universal distance in
nature, not unlike the postulate about the speed of light as the
upper limit on the speed of signal propagation.

What seems remarkable in this scheme of things is the nature of
the signatures and the emergence of two times. One of the latter
is the local mode, a clock, represented by $ t $ and the other
mode is a ``global'' one represented by the volume of the space-time
filling brane $ \Omega $. For more details related to this
Fock-Stuckelberg-type parameter see [3]. We must emphasize that one 
must
not confuse these global and local time modes with the two modes of 
time in
other branches of science [13].

Another immediate application of this theory is that one may
consider ``strings'' and ``branes'' in C-spaces as a unifying
description of $ all$ branes of different dimensionality. As we
have already indicated, since spinors are left/right ideals of a
Clifford algebra, a supersymmetry is then naturally incorporated
into this approach as well. In particular, one can have world
volume and target space supersymmetry $simultaneously$ [17].
We hope that the $C$-space ``strings'' and ``branes'' may lead us
towards discovering the physical foundations of string and
M-theory.

\bigskip
\centerline {\bf 1. 3 The Generalized String/Brane Uncertainty 
Relations }
\bigskip

Below we will review how the minimal length string uncertainty 
relations
can be obtained from C-spaces
[2]. The norm of a momentum poly-vector was defined:

$$ P^2 = \pi^2 + p_\mu p^\mu + p_{\mu\nu} p^{\mu\nu} + p_{\mu\nu\rho}
p^{ \mu\nu\rho} + .... = M^2 \eqno (12 ) $$ Nottale has given
convincing arguments why the notion of $dimension$ is resolution
dependent, and at the Planck scale, the minimum attainable
distance, the dimension becomes singular, that is blows-up. If we
take the dimension at the Planck scale to be infinity, then the
norm $ P^2$ will involve an infinite number of terms since the
degree of a Clifford algebra in $ D$-dim is $ 2^D$. It is
precisely this infinite series expansion which will reproduce
$all$ the different forms of the Casimir invariant masses
appearing in kappa-deformed Poincare algebras [11,12].

It was discussed recently why there is an infinity of possible
values of the Casimirs invariant $ M^2$ due to an infinite choice
of possible bases. The parameter $ \kappa$ is taken to be equal to
the inverse of the Planck scale. The classical Poincare algebra is
retrieved when $\Lambda = 0$. The kappa-deformed Poincare algebra
does $not$ act in classical Minkwoski spacetime. It acts in a
quantum-deformed spacetime. We conjecture that the natural
deformation of Minkowski spacetime is given by C-space.

The way to generate all the different forms of the Casimirs $ M^2$
is by ``projecting down'' from the $ 2^D$-dim Clifford algebra to
$ D$-dim. One simply ``slices'' the $ 2^D$-dim mass-shell
hyper-surface in C-space by a $D$-dim one. This is achieved by
imposing the following constraints on the holographic components
of the polyvector-momentum. In doing so one is explicitly
$breaking$ the poly-dimensional covariance and for this reason one
can obtain an infinity of possible choices for the Casimirs $
M^2$.

To demonstrate this, we impose the following constraints:

$$ p_{\mu\nu} p^{\mu\nu} = a_2 (p_\mu p^\mu)^2 = a_2 p^4. ~~~
p_{\mu\nu\rho} p^{\mu\nu\rho} = a_3 (p_\mu p^\mu)^3 = a_3 p^6 . ~~~
...... \eqno ( 13 ) $$

Upon doing so the norm of the poly-momentum becomes:

$$ P^2 = \sum_n a_n p^{2n} = M^2 ( 1, a_2, a_3, ..., a_n, ...) \eqno
(14 )
$$
Therefore, by a judicious choice of the coefficients $ a_n $,
and by reinserting the suitable powers of the Planck scale, which
have to be there in order to combine objects of different
dimensions, one can reproduce $all$ the possible Casimirs in the
form:

$$ M^2 = m^2 [ f ( \Lambda m /\hbar )] ^2 . ~~~ m^2 \equiv p_\mu p^\mu 
=
p^2.
\eqno ( 15 ) $$ where the functions $ f ( \Lambda m/\hbar )$
are the
$scaling$ functions with the property that when $ \Lambda = 0$
then $ f \rightarrow 1 $.
\bigskip

To illustrate the relevance of poly-vectors, we will summarize our
derivation of the minimal length string uncertainty relations [2].
Because of the existence of the extra holographic variables one cannot
naively impose $ [ x, p] = i\hbar $ due to the effects of the other
components. The units of $[ x_{\mu\nu}, p^{\mu\nu} ] $ are of
$\hbar^2 $ and of higher powers of $\hbar$ for the other
commutators. To achieve covariance in C-space which reshuffles
objects of different dimensionality, the effective Planck
constant in C-space should be given by a sum of powers of $
\hbar$.

This is not surprising. Classical C-space contains the Planck
scale, which itself depends on $\hbar$. This implies that already
at the classical level, C-space contains the seeds of the quantum
space. At the next level of quantization, we have an effective
$\hbar$ that comprises all the powers of $\hbar$ induced by the
commutators involving $all$ the holographic variables. In general
one must write down the commutation relations in terms of
polyvector- valued quantities. In particular, the Planck constant
will now be a Clifford number, a polyvector with multiple
components. This will be the subject of section {\bf II }.

The simplest way to infer the effects of the holographic
coordinates of C-space on the commutation relations is by working
with the effective $ \hbar$ that apperas in the $nonlinear$ de Broglie
dispersion relation. The mass-shell condition in C-space, after 
imposing the
constraints among the holographic
components, yields an effective mass $ M = m f ( \Lambda m/\hbar ) $.
The generalized De Broglie relations, which are $no$ longer
linear, are [2]:

$$| P_{effective} | = | p | f ( \Lambda m/\hbar ) =
\hbar_{effective}
| k| .
~~~ \hbar_{effective} = \hbar f ( \Lambda m/\hbar ) = $$
$$\hbar \sum a_n (\Lambda m/\hbar )^{ 2n} . ~~~ m^2 = p^2 = p_\mu 
p^\mu =
( \hbar
k)^2. \eqno ( 16 ) $$

Using the effective $ \hbar_{eff}$, the well known relation based
on the Schwartz inequality and the fact that $ | z | \ge | Im z |
$ we obtain:

$$ \Delta x^i \Delta p^j \ge { 1 \over 2 } | < [ x^i, p^j ] > | .
~~~[ x^i, p^j ] = i \hbar_{eff} \delta^{ij}. \eqno ( 17 ) $$

Using the relations

$$< p^2 > \ge ( \Delta p )^2 .~~~ < p^4 > \ge ( \Delta p )^4 . .....
\eqno (18 ) $$

and the series expansion of the effective $ \hbar_{eff}$, we get
for each component (we omit indices for simplicity):

$$ \Delta x \Delta p \ge { 1\over 2 } \hbar + { a \Lambda^2 \over 2
\hbar } ( \Delta p)^2 +............\eqno ( 19 ) $$

This yields the minimal length string uncertainty relations:

$$ \Delta x \ge { \hbar \over 2 \Delta p } + { a \Lambda^2 \over 2
\hbar } \Delta p .....\eqno (20 ) $$

By replacing lengths by times and momenta by energy one reproduces the
minimal Planck time uncertainty relations. One could include $all$ the
terms in the series expansion and derive a generalized string/brane
uncertainty relation which still retains the minimal
length condition, of the order of the Planck scale [2]. For example, if 
one
chooses the same value for all the coefficients in the Taylor 
expansion, an
$isotropy$ condition in C-spaces is selected where all directions have 
equal
weight, the full blown uncertianty relation due to $all$ branes is 
given by
[2]:

$$ \Delta x \ge \sqrt {2 } \Lambda { e^{ (\Delta z)^2 /4 } \over (
\Delta z )^2 } \sqrt { sinh~ [ ( \Delta z )^2 / 2 ] }.
\eqno ( 21) $$
where $ \Delta z = \Lambda \Delta k $ and $ k = \sqrt { k_\mu k^\mu } $ 
and
we took all the coefficients of the Taylor expansion to be equal to 
unity.
This relation also obeys the minimal length condition [2] of the order 
$
1.2426 ~ \Lambda $. Uncertainty relations for a particular $p$-brane
(for a specific value of $p$) has been given by [20]. Relations given by 
eq-(21) are due to the contribution of $ all$
values of $ p$. In the limit that $ \Lambda $ goes to zero we recover 
the
standard Heisenberg uncertainty relations.

The Physical interpretation of these uncertainty relations follow
from the extended relativity principle. As we boost the string to
higher $transPlanckian$ energies part of the energy will $always$ be
invested into the string's potential energy, increasing its length 
in bits
of
Planck scale sizes so that the original string will decompose into two,
three, four....strings of Planck sizes carrying units of Planck 
momentum;
i.e. the notion of $a$ single
particle/string loses its meaning beyond that point. This reminds one 
of 
ordinary relativity,
where boosting a massive particle to higher energy increases its
speed while a part of the energy is also invested into increasing
its mass. In this process the speed of light remains the maximum
attainable speed (it takes an infinite energy to do so) and in
our scheme the Planck scale is never surpassed. The effects of a
minimal length can be clearly seen in Finsler geometries having
both a maximum four acceleration $ c^2 / \Lambda$ (maximum tidal
forces) and a maximum speed [21]. The Riemannian limit is
reached when the maximum four acceleration goes to infinity; i.e. The
Finsler geometry ``collapses'' to a Riemannian one.

It is straigthforwad now to derive the maximum Planck temperature 
condition
[14] and the Thermodynamic Uncertainty relations [19]. Based on the old
known results of Euclidean QFT, we simply identify the inverse 
temperature $
1/ T$ with the period of the Euclideanized temporal coordinate $ x_0$.
By
using the simple correspondence in
eqs-(20):

$$ x_0 \rightarrow 1/T. ~~~ \hbar \rightarrow k_B.~~~ \Delta E 
\rightarrow
\Delta U $$

we will recover the maximum Planck temperature Uncertainty relations
[19]:

$$ \Delta ( 1 / T ) \ge { k_B \over 2 \Delta U } + { a \over 2 k_B
T_P
^2 }
\Delta U \eqno ( 22 ) $$
in terms of the temperature $ T$, the internal enegy $ U$ and the 
Boltzman
constant $ k_B$.
The Planck temperature $ T_P$ is defined by $ M_P / k_B $ in units of $ 
c =
1 $.

\bigskip

\centerline{\bf II. Weyl-Heisenberg Algebra in C-spaces }

\bigskip

A straightforward procedure to visualize the C-space 
algebraic-geometric
structure can be achieved by recalling that a realization of the basis
elements $ E_A $ exists in terms of Dirac matrices, and their suitable
antisymmetrized products of matrices, until saturating the 
dimensionality
of spacetime. A Clifford algebra in $ D$ dimenssion has $ 2^D$ basis
elements including the unit element. For $ D = 2n $ one has a basis of
$2^D$ matrix-elements where each matrix-element is given by a $ 2^n 
\times
2^n$ Dirac matrix, for example.

Due to the noncommutative nature of the basis vectors of the Clifford
algebra one has:

$$ [ E_A, E_B ] = F^M_{AB} E_M = F_{ABM} E^M. ~~~~
E_A = \{ I; \gamma_\mu; \gamma_\mu \wedge \gamma_\nu; \gamma_\mu 
\wedge
\gamma_\nu \wedge \gamma_\rho;.....\}. \eqno (23 a) $$
the quantities $ F^M _{AB}$ play a similar role as the structure 
constants
in ordinary Lie algebras. A commutator of two matrices is itself a 
matrix,
which in turn, can be expanded in a
suitable matrix basis due to the Clifford algebraic (vector space)
structure inherent in C-spaces. The commutator algebra obeys the Jacobi
identities, the Liebnitz rule of derivations and the antisymmetry
properties.

The Clifford geometric product of two basis elements is defined in 
terms of
a ``scalar'' (symmetric) and ``outer'' product (antisymmetric)
respectively:

$$ E_A E_B = {1 \over 2 } \{ E_A, E_B \} + {1 \over 2 } [ E_A, 
E_B ].
\eqno ( 23 b) $$

In general the geometric product of two multivectors $E^A, E^B $
of ranks $ r, s $, respectively, is given by an aggregate of 
multivectors of the form:

$$ E^A E^B = < E^A E^B>_{r + s }, ~ <E^A E^B >_{r+s - 2 }, ~
<E^A E^B >_{r + s - 4 },......<E^A E^B>_{ |r -s| } $$

The first term of rank $ r+s $ is the wedge product $ E^A \wedge E^B$ 
and 
the last term of rank $ |r - s| $ is the $dot$ product $ E^A.E^B$ which 
is 
obtained by a $contraction$ of indices and must not be confused with 
the 
scalar part of $ E^A E^B$ unless $ r = s $. In general, the scalar 
product 
among two equal-rank multivectors $ r = s $ cannot longer be written in 
terms of the anticommutator $ \{ E^A, E^B \} $ except in the case when
$ r = s = 1 $: $ \{ \gamma^\mu, \gamma^\nu \} = 2 g^{\mu\nu} 1. $
However, for equal-rank multivectors, the scalar part
$ <E_A E_B>_{0} = G_{AB} I $ where $ I $ is the unit element of the 
Clifford algebra and $ G_{AB} $ is the C-space metric.

To begin with we will write for the putative Weyl-Heisenberg algebra:

$$ [ X^A, P^B ] = H^0 G^{AB} - H^{{\hat C} } F^{AB}_ {{ \hat C}}.
\eqno (
24 ) $$
where $ H^{ {\hat C } }, H^0 $ are themselves the components of a
polyvector-valued C-space generalization of Planck's constant
$ H = H^C E_C$. The $ {\hat C}$ multi-index runs over all the basis 
elements
$except$ the unit element $ H^0 I $. Later on we will see that this
relation needs to be modified by the addition of crucial terms 
involving the
$ X, P$ variables.

First of all we must keep track of the correct units. We will treat the 
$ P
$ and $ X$ exactly on the same footing. For this reason we will 
$scale$ all
the basis elements by judicious powers of
$ \sqrt {\hbar} $:

$$ \gamma^\mu \rightarrow \sqrt {\hbar} \gamma^\mu. ~~~~
\gamma^\mu \wedge \gamma^\nu \rightarrow \hbar \gamma^\mu \wedge
\gamma^\nu;.....\eqno ( 25 a ) $$
and

$$ \gamma_\mu \rightarrow {\hbar}^{ - 1/2 } \gamma_\mu. ~~~~
\gamma_ \mu \wedge \gamma_\nu \rightarrow \hbar^{ - 1 }
\gamma_ \mu \wedge \gamma_ \nu;.....\eqno ( 25b) $$

The units of $ X^A $ and $ P^B$ are taken to be $ {\hbar}^{ r_A /2 }$ 
and
$ {\hbar}^{ r_B/2 }$ respectively where $ r_A, r_B$ are the ranks of 
the
antisymmetric tensor components $ X^A, P^B $ of the polyvectors $ X, P$
respectively.

The scaling of the commutator is:

$$ [ E_A, E_B ] = F^M_{AB} E_M \rightarrow \hbar^{- r_A/2 - r_B/2 
}[
E_A, E_B ] =
( \hbar^{- r_A/2 - r_B/2 } F^M_{AB} ) ~ E_M. \eqno ( 26 ) $$
hence, we will absorb the powers of $ \hbar$ appropriately in the 
structure
constants as indicated above. Upper indices carry positive powers of 
$\sqrt
{\hbar} $ whereas lower indices carry negative powers. Notice that 
there
are $no $ powers of $ \sqrt {\hbar} $ associated with the index $ M$ 
above,
only w.r.t the two $ A B $ indices.

Since we are scaling the basis vectors $ E_A $ this means that we are
choosing the $ H^{ { \hat C } }, H^0 $ to be dimensionless. In 
this 
fashion we automatically obtain quantities with the correct units. For 
example:
$ [ x^\mu, p^\nu ]$ will contain a single power of $\hbar$ due to the 
two
factors of
$\hbar^{1/2}$ appearing in the $ G^{\mu \nu } $ as a result of the 
scaling
of the $ \gamma^\mu $ and $ \gamma^\nu $. Their anticommutator yields 
the $
G^{\mu\nu} $ (after the saling takes place). One obtains identical 
results 
with the
other holographic components of the polyvectors. The commutator of $ [
x^{\mu \nu }, p^{\rho \tau} ] $ will automatically have the correct $
\hbar^2 $ power, etc...... Identical results follow for the
$ H^{{\hat C}} F^{AB}_{{\hat C}} $ terms as well.

If one uses these putative Weyl-Heisenberg algebra relations in the 
Jacobi
identities for the set of variables $ X^A, P^B, X^C $ ordinary
commutatitivity of the coordinates will be maintained,
$ [ X^A, X^C ] = 0 $. However this is not longer the case in the
full-fledged algebra as we shall see next.

A direct evaluation of the commutator of two $polyvectors $ in terms of 
the Clifford-valued Planck constant $ H= H_M E^M$ is:

$$ [ X, P ] = H = [ X_A E^A, P_B E^B ] = H = H^C E_C = 
[ X_A, P_B ] E^A E^B + P^M X^N [ E_M, E_N ] = $$
$$ [ X_A, P_B ] E^A E^B + P^M X^N F_{MN}^C E_C 
 \eqno ( 27 ) $$
the $ E^A E^B$ terms in the r.h.s of (27) can be reshuffled to the 
l.h.s 
by means of writing the inverse of the geometric product as:

$$ ( E_A E_B )^{ - 1 } = E_B^{ -1} E_A^{ -1 } = E^B E^A. \eqno ( 28 ) 
$$
and this allows us to write the Weyl-Heisenberg algebra in C-spaces in 
terms 
of the $scalar$ part of the $triple$ geometric product $ < E_C E_B E_A 
> $ 
as:

$$ [ X_A, P_B ] = ( H^C + P^M X^N F^C_{ MN}) < E_C E_B E_A >_0.
\eqno ( 29 ) $$
where $ <E_C E_B E_A >_0 \equiv \Omega_{CBA} $ is the scalar part of 
the 
geometric triple product. Eq-(29) is the fundamental result of this 
work.

Inspired on this result (29), if one wishes to write the 
Weyl-Heisenberg 
algebras in terms of $ G^{AB} $ and the structure constants $F^{AB}_M, 
K^{AB}_M $ of the commutators and anticommutators, respectively,
$ [E^A, E^B ] = F^{AB}_C E^C $, $ \{ E^A, E^B \} = K^{AB}_C E^C $, 
the Weyl-Heisenberg algebra reads:

$$ [ X_A, P_B ] = H_0 G_{AB} + H_{{\hat C}}
[ F_{BA}^{{\hat C}} + K_{AB}^{{ \hat C }} ] + $$

$$ P_M X_N F^{MN}_C [ F_{BA}^C + K_{AB}^C ]. $$
Once again the ${\hat C}$ index
in (27b) runs over all multi-indices of the
Clifford algebra except the unit element.

The Weyl-Heisenberg algebra can be written compactly in the 'spin'
plus 'orbital' angular momentum form:

$$ [ X_A, P_B ] = H_{ AB } + J_{ AB}. \eqno ( 30 ) $$

with the standard Planck constant-like terms of the form:

$$ H_{ AB} = H_0 G_{AB} + H_{{\hat C }}
[ F_{BA}^{{\hat C }} + K_{AB}^{{ \hat C}} ]. \eqno ( 31) $$
Notice the $mixed$ symmetry of this expression, a symmetric plus
antiymmetric piece in $ A, B $. Had one had commuting basis elements, 
like
in ordinary spacetime, and a scalar component for $ H = H_0 $ one would 
have 
had the standard Weyl-Heisenberg
algebra $ [ X^A, P^B ] = H^0 G^{AB} $. Since the powers of $ \hbar$ are
absorbed by the metric $ G^{ AB} $ this implies that $ H^0 = i $.
The extra term in (30) is the analog of the orbital angular momentum 
in 
C-spaces
given by:

$$ J_{ AB } = P_M X_N F^{ MN}_C < E^C E_B E_A>_0
\equiv J_C <E^C E_B E_A >_0 = J_C \Omega^C_{ AB }. \eqno ( 32 ) $$
it also has mixed symmetry in the indices $ A, B $.

The elements $ H^{ AB} $ involving the components of the 
polyvector-valued
Planck constant resemble the familiar quaternionic and octonionic 
expansions
of a quaternion and octonion in terms of their components. This 
indicates
that QM in C-spaces may be intrinsically linked with Quaternionic and
Octonionic QM [23].

An immediate consequence of the C-space Weyl-Heisenberg algebra is that 
it
$induces$ automatically a Noncommutative Geometric structure in the $ 
X^A $
coordinates. To satisfy the Bianchi identities among the triples $ 
X^A,
P^B, X^C $ and $ X^A, X^B, X^C $ it is fairly clear that the 
coordinates $
cannot $ commute due to the explicit $ X, P$ terms in the modified
Weyl-Heisenberg algebra. Hence, the Jacobi identities require:

$$ [ X^A, X^B ] = \Sigma^{ AB}. ~~~ [X^C, \Sigma^{AB} ] = 0. \eqno ( 
33 )
$$

where $ \Sigma^{ AB} $ is a tensor-like operator-valued object in 
C-space
that does not
destroy C-space Lorentz invariance and which is implicitly defined by 
the
Jacobi identities. Suitable powers of the Planck scale are absorbed in 
the
defining relations for
$ \Sigma^{ AB} $ in order to match units. Since the Planck scale is a
C-space invariant one will maintain C-space Lorentz invariance. To 
evaluate
explictly the expression for $ \Sigma^{AB} $ will be the subject of 
future
investigation. It is nontrivial even if we set in $flat$ C-spaces:
$ [ P^A, P^B ] = 0 $.

Another important consequence is that we $cannot$ represent naively
the operators by the old QM prescriptions:

$$ X^A \rightarrow ( H^{ AB} + J^{AB} ) { \partial \over \partial P^B 
}.
\eqno ( 34 a ) $$

$$ P^A \rightarrow( H^{ AB} + J^{ AB} ) { \partial \over \partial X^B 
}.
\eqno ( 34 b ) $$

These naive representations of the $ X, P $ operators in the 
Weyl-Heisenberg
algebra do $ not $ longer hold due to the explicit $ X, P$ 
dependence of
the C-space angular momentum $ J^{ AB} $ in the Weyl-Heisenberg 
algebra.

Using the effective $ \hbar ( p^2 ) $, where $ p^2 = p_\mu p^\mu $, [2]
we could still represent the position operator in terms of the momentum
variables:

$$ x_i \rightarrow i \hbar_{ effective} ( p^2 ) { \partial \over 
\partial
p^i }. \eqno ( 35 a ) $$

but $no$ longer we may write that:

$$ p_i \rightarrow - i \hbar_{ effective} ( p^2 ) { \partial \over 
\partial
x^i }.\eqno ( 35 b ) $$
otherwise one would not have been able to satisfy the Weyl-Heisemerg
relation:

$$ [ x^i, p^j ] = i \hbar_{eff} ( p^2) \delta^{ij}. \eqno ( 36 ) $$
assuming a flat spacetime $ [ p^i, p^j ] = 0 $.
Hence, the symmetry between $ x, p $ is $broken$ already in these 
cases
where one works with a modified Weyl-Heisenberg algebra using an 
effective
$\hbar ( p^2 ) $ with $ p^2 = p_\mu p^\mu $.

Using other effective
matrix valued $\hbar_{ij} $ that depend on $ \vec p = p^i $ and on
products
like $ p^i p^j$ happen to $break$ Lorentz invariance explicitly despite
maintaining rotational symmetry [21, 25]. The fact that Lorentz
invariance is broken is not surprising since these models are based on
kappa-deformed Poincare symmetries [11, 12]. We haves hown how one can
break C-space Lorentz invariance to obtain a kappa-deformed effective
Lorentz transformations which leave the Planck
scale invariant [2, 16].

A length scale was introduced by $hand$ by Snyder [22]
when he wrote down the commutation relations for the four spacetime
coordinates based on a
$5$-dimensional spacetime, whose $ fifth$ dimension was $discrete$ in 
units
of length $ l $:

$$ [ x^\mu, x^\nu ] = l^2 [ J^{5\mu}, J^{5\nu} ] =
i l^2 J^{ \mu \nu}$$
where $ J^{ \mu \nu } $ is an angular
momentum in $ four$ dimensions and the four spacetime coordinates $ 
x^\mu $
(divided by $ l $) are identified with the components of the angular
momentum which contain the
$ fifth$ direction. Lorentz
invariance is maintained in the four-dimensions by construction.
The origin of the scale $ l$ is due to the discrete fifth dimension.
The authors in [11,12 21] have related this scale to the deformation
parameter of kappa-deformed Poincare algebras $ l = 1 / \kappa = 
\Lambda $,
with the fundamental difference
that the four dim Lorentz invariance is $ broken$, only rotational 
symmetry
is conserved.

The advantage of C-spaces is that one does not need to introduce 
$ad-hoc$
this angular momentum type commutators for the four spacetime 
coordinates,
by recurring to an extra discrete dimension of size $ l $. The
Weyl-Heisenberg and $ X^A $ coordinate algebras in C-space are indeed
compatible with the C-space Lorentz invariance $without$ introducing 
extra
dimensions.
The Planck scale is a true invariant of C-space. In [13] we have shown
that the Conformal algebra in four dimensions $ SO(4,2) $ does $not$
require the six-dimensional
interpretation associated with the Anti de Sitter group. Instead, it 
can be
obtained directly from the Clifford algebra of four dim spacetime.

It is fairly clear why C-space QM differs from the ordinary QM in many
aspects.
To start it is already a Noncommutative QM since it involves a
Noncommutative Geometric structure for the $ X$ coordinates. The main
task
in the near future will be to construct a QFT
in C-spaces, in particular, to use Quantum Clifford Algebras and 
Braided
Hopf Quantum Cliiford algebras to study q-deformations of C-spaces 
[15].
A Moyal-like star product construction deserves further study as well.
Since C-spaces involves the physics of all $ p$-loops it is warranted 
to use
methods of multisymplectic geometry (mechanics) since phase spaces in
C-spaces involve antisymmetric tensors of arbitrary rank. Nambu-Poison 
QM
seems to be the most appropriate one to study C-space QM. In 
particular the
use of the Zariski star product deformations vz the Moyal one [24] will 
be
welcome.

To finalize we discuss how one would take $contractions$ of the
Weyl-Heisenberg algebra to obtain an effective matrix valued $ 
\hbar_{ij}
$, or for that matter, to generate a single effective Planck constant 
in the
form of $ \hbar \delta_{ij} $. The most natural candidate is to take 
the
norm-squared as an effective $ \hbar$:

$$ H^{AB} H_{ AB} + J^{AB} J_{ AB} + H^{AB} J_{ AB} + J^{AB} H_{AB}.
\eqno ( 37 ) $$

C-space polydimensional invariance can be broken by imposing similar 
type of
constraints like we had in section {\bf 1 }, relating the holographic 
norms
of polyvectors to the powers of ordinary vector norms:

$$ J^{AB} J_{AB} = \sum_n a_n (J/\hbar )^{ 2n }. ~~~ J^2 = J_{\mu \nu 
}
J^{\mu
\nu
}. ~~~
J^{\mu \nu } = p^\mu x^\nu - p^\nu x^\mu. ~~~ \eqno ( 38 ) $$
and in this way the norm-squared (37) reduces to:

$$ (\hbar_{eff} / \hbar )^2 = H^{AB} H_{ AB} + H^{AB} J_{AB} +
J^{AB} H_{AB} + \sum_n a_n (J/\hbar)^{ 2n } \eqno
( 40 ) $$

Eq-(40) for the effective $ \hbar (p^2) $ is far more general than 
the 
ones discussed in the previous section. In particular, the last terms 
of (40) do contain the required terms $ \sum p^{2n} $ in the effective
$\hbar$.
This can be understood after relating the angular momentum $ J$ to the 
$ m^2 = p^2 $ of the center of mass coordinates. In the ordinary mechanics of 
a 
rigid top there are two Casimirs,
the angular momentum $ J^2 $ and the Energy given in terms of the 
angular
momentum, in the case of a $symmetric$ top, as $ E = J^2 / 2 {\cal 
I} $
where $ {\cal I} $ is the moment of inertia of the symmetric top.

If one assumes a similar relationship among Energy and angular momentum 
in
the relativistic case, one will have the familiar Regge-type of 
relation
asociated with the on-shell string specrum:
$ J \sim (\alpha)' ~ m^2 $, where on shell, $ m^2 = p^2 $ and $ 
(\alpha)'
$ is the inverse string tension, of the order of
$ \Lambda^2 $. In this case one will be able to match the results of
section {\bf 1 }, involving an effective $ \hbar_{eff} ( p^2 ) $, for 
on-shell values $ p^2 = m^2 $, with the last terms of (40) after 
breaking 
the C-space polydimensional invariance in (38) and using the desired 
Regge 
relation.

Concluding: Quantization in C-spaces contains a very rich 
Noncommutative
structure from which many old results can be derived after breaking the
C-space Lorentz invariance. There is no need to introduce ad-hoc 
nontrivial
commutation relations for the spacetime coordinates. These are induced 
from
the mere quantization process. No extra discrete fifth dimension is
required to introduce a length scale.
C-space Relativity already has a natural invrariant minimum Planck 
scale by
definition.

\bigskip
\centerline{\bf Acknowledgements }
\bigskip

We are indebted to Matej Pavsic and Alex Granik for many conversations
which led to this work. To Jorge Mahecha for his help in preparing the
manuscript and to Dr.  Alfred Schoeller and family for their warm
Mistelbachian hospitality without which this work would not have been
possible.

\newpage

\centerline{\bf References }
\bigskip

1- L. Nottale: ``La Relativite dans tous ses Etats''. Hachette
Lit. Paris 1999.

``Fractal Spacetime and Microphysics, Towards Scale Relativity''. World
Scientific. Singapore, 1992.

2- C. Castro: Chaos, Solitons and Fractals {\bf 11} (2000) 1663.
hep-th/0001134
Foundations of Physics {\bf 30} (2000) 1301. hep-th/0001023.
Chaos, Solitons and Fractals {\bf 11} (2000) 1721. hep-th/9912113. 
Chaos,
Solitons and Fractals {\bf 12} (2001) 1585. physics/0011040. ``The
Programs
of the Extended Relativity in C-spaces, towards the physical 
foundations of
string theory'' hep-th/0205065. ``The search for the origins of M 
theory
....'' hep-th/9809102

3- M.Pavsic: ``The Landscape of Theoretical Physics'': A Global View
Kluwer,
Dordrecht 1993.
``Clifford algebra based polydimensional Relativity and Relativistic
Dynamics'' Talk presented at the IARD Conference in Tel Aviv, June 
2000.
Foundations of Physics {\bf 31} (2001) 1185. hep-th/0011216. Phys. Let 
{\bf
A 242} (1998) 187. Nuovo Cimento {\bf A 110} (1997) 369.

4- J. Fanchi ``Parametrized Relativistic Quantum Theory''. Kluwer,
Dordrecht 1993.

5- W. Pezzaglia: ``Physical Applications of a Generalized
Geometric Calculus'' gr-qc/9710027.

6- D. Hestenes: ``Spacetime Algebra'' Gordon and Breach, New York, 1996.

D. Hestens and G. Sobcyk: ``Clifford Algebra to Geometric
Calculus'' D. Reidel Publishing Company, Dordrecht, 1984.

7- ``Clifford Algebras and their applications in Mathematical 
Physics''
Vol 1: Algebras and Physics. eds by R. Ablamowicz, B. Fauser.
Vol 2: Clifford analysis. eds by J. Ryan, W. Sprosig Birkhauser, Boston
2000.

8- P. Lounesto: ``Clifford Algebras and Spinors''. Cambridge
University Press. 1997.

9- S. Ansoldi, A. Aurilia, E. Spallucci: Chaos, Solitons and
Fractals {\bf 10} (2-3) (1999).

10- S. Ansoldi, A. Aurilia, C. Castro, E. Spallucci: Phys. Rev.
{\bf D 64 } 026003 (2001) hep-th/0105027.

11- J. Lukierski, A. Nowicki:`` Double Special Relativity versus
kappa-deformed Relativistic dynamics'' hep-th/0203065.

J. Lukierski, V. Lyakhovsky, M. Mozrzymas: ``kappa-deformations
of D = 4 Weyl and conformal symmetries'' hep-th/0203182.

S.Majid, H. Ruegg: Phys. Let {\bf B 334} (1994) 348.

J. Lukierski, H. Ruegg, W. Zakrzewski: Ann. Phys. {\bf 243 }
(1995) 90.

J. Lukierski, A. Nowicki, H, Ruegg, V. Tolstoy: Phys. Lett {\bf B
264} (1991) 331.

12- J. Kowalski-Glikman, S. Nowak: ``Noncommutative spacetime of Double
Special Relativity'' hep-th/0204245

13- C. Castro, M. Pavsic: ``Higher Derivative Gravity and Torsion from the
Geometry of C-spaces'' hep-th/0110079. To appear in Phys.  Lets B.

C. Castro, M. Pavsic: ``The Clifford algebra of Spacetime and the
Conformal Group'' hep-th/0203194.

14- C. Castro, A. Granik: ``Extended Scale Relativity, p-loop harmonic
oscillatorand logarithmic corrections to the black hole entropy''
physics/0009088.

C. Castro: Jour. of Entropy {\bf 3} (2001) 12-26.

15- B. Fauser: ``A treatise on Quantum Clifford Algbras''
math.QA/0202059

Z. Osiewicz: ``Clifford Hopf Algebra and bi-universal Hopf
algebra'' q-qlg/9709016.

C. Blochmann ``Spin representations of the Poincare Algebra''
Ph. D Thesis math.QA/0110029.

16- C. Castro, A. Granik: ``Planck scale Relativity and variable
fine structure from C-space'': To appear.

17- A. Aurilia, C. Castro, M. Pavsic, E. Spallucci: To appear.

18- H. Brandt: Chaos, Solitons and Fractals {\bf 10} (2-3)
(1999) 267.

19- A. Margolin, A. Tregubovich: ``Generalized uncertainty
relations.....and thermodynamics from a uniform point of view''
gr-qc/0204078.

20- S. Ansoldi, A. Aurilia, E. Spallucci: ``Fuzzy dimensions and 
p-brane
uncertainty relation''
hep-th/ 0205028. To appear in Class. Quan. Gravity.

21- M. Maggiore: ``The Atick-Witten free energy, closed tachyon
condensation and deformed Poincare algebra'' hep-th/0205014.

22- S. Snyder: Phys. Rev. {\bf D } 71 (1947) 38. Phys. Rev. {\bf D } 71
(1947) 78.

23- S. Adler: ``Quaternionic QM and QFT'' Oxford Univ. Press. New York,
1995.

24- G. Dito, M. flato, D. Sternheimer, L. Takhtajan: Comm. Math. Phys.
{\bf 183 } (1997) 1-22.

25- A. Kempf, G. Mangano, R. Mann: Phys. Rev {\bf D 52 } (1995) 1108.
hep-th/9412167.

\end{document}